*Quantum Geometrodynamics Revived*
*II. Hilbert Space of Positive Definite Metrics*

Thorsten Lang[*,1] and Susanne Schander[†,2]

[1]Institute for Quantum Gravity, FAU Erlangen – Nürnberg, Staudtstraße 7/B2, 91058 Erlangen, Germany
[2]Perimeter Institute, 31 Caroline St N, Waterloo, ON N2L 2Y5, Canada

June 25, 2023

*Abstract*

This paper represents the second in a series of works aimed at reinvigorating the quantum geometrodynamics program. Our approach introduces a lattice regularization of the hypersurface deformation algebra, such that each lattice site carries a set of canonical variables given by the components of the spatial metric and the corresponding conjugate momenta. In order to quantize this theory, we describe a representation of the canonical commutation relations that enforces the positivity of the operators $\hat{q}_{ab}s^a s^b$ for all choices of *s*. Moreover, symmetry of $\hat{q}_{ab}$ and $\hat{p}^{ab}$ is ensured. This reflects the physical requirement that the spatial metric should be a positive definite, symmetric tensor. To achieve this end, we resort to the Cholesky decomposition of the spatial metric into upper triangular matrices with positive diagonal entries. Moreover, our Hilbert space also carries a representation of the vielbein fields and naturally separates the physical and gauge degrees of freedom. Finally, we introduce a generalization of the Weyl quantization for our representation. We want to emphasize that our proposed methodology is amenable to applications in other fields of physics, particularly in scenarios where the configuration space is restricted by complicated relationships among the degrees of freedom.

*Contents*

[*]`thorsten.lang@fau.de`
[†]`sschander@perimeterinstitute.ca`

# 1 Introduction

Despite the existence of several promising approaches to quantum gravity, the quest for a viable theory that successfully unites general relativity with the principles of quantum mechanics, and resolves all associated challenges, remains an elusive endeavor.

One of the earliest attempts to quantize gravity was the approach of Quantum Geometrodynamics that employs a 3 + 1–split of spacetime (cf. Arnowitt et al. 2008; DeWitt 1967). In this framework, the spatial metric field $q_{ab}(x)$, with and its conjugate momentum $p^{ab}(x)$ constitue the basic canonical variables of the theory. At the classical level, $q_{ab}$ is a symmetric and positive definite tensor field. The last requirement ensures that the spacetime metric $g$, which is built from spatial metric (in addition the lapse function and shift vector field), has a Lorentzian signature, which amounts to the implementation of relativistic causality within the theory.

In order to quantize this theory in accordance with Dirac's approach to the quantization of constrained systems, it is necessary to represent the constraints as quantum operators on a suitable Hilbert space. However, this poses several challenges (Isham 1991; Kiefer 2007). For instance, it remains an open question as to how construct a well–defined (kinematical) Hilbert space for the representation of the quantum constraints which are highly non–polynomial functions of the canonical variables. Another challenge concerns the realization of a positive definite spatial metric in the quantum theory. In fact, if the matrix elements of the metric and their conjugate momenta were quantized in the standard way, i.e., by representing them by multiplication and derivative operators, then states with support on non–positive definite matrices were not forbidden. One well–known way to overcome this is expressing the metric in terms of triad fields. While the introduction of triads is necessary for defining spinor fields on the manifold, the standard representation still ensures only positive semi–definiteness and fails to fix the orientation of the triads. Another proposal comes from affine gravity where one ensures positive definiteness by assuming a different set of commutation relations for the canonical variables.

We refer to (Lang and Schander 2023a,c) and references therein for further details on these issues and others. In this series of papers, of which this is the third one, we address several of these challenges and propose solutions in order to reinvogarate the quantum geometrodynamics approach.

The present work centers on the development of a rigorous Hilbert space formulation for quantum geometrodynamics on the lattice. As detailed in (Lang and Schander 2023a), we first discretize classical geometrodynamics, reducing the infinite degrees of freedom to a finite set of $n(n+1)/2$ components per lattice point, corresponding to each component of the symmetric spatial metric tensor in $n$ dimensions. This finite–dimensional approach facilitates the quantization of the theory using standard quantum mechanics techniques. The primary objective of this paper is to construct a well–defined Hilbert space for both individual lattice sites and the overall system. To achieve this, we introduce a novel representation of the ususal commutation relations which manifestly ensures the positive definiteness of the spatial metric tensor.

Specifically, we begin our endeavor by employing the Cholesky decomposition (Benoît 1924; Golub and Van Loan 2013) to represent the spatial metric. More precisely, the Cholesky decomposition enables us to express any positive definite matrix as the product of an upper triangular matrix with positive diagonal elements and its transpose. This subset of upper triangular matrices having positive diagonal elements constitutes a Lie group, which consequently allows us to endow the space of positive definite symmetric matrices with a Lie group structure. In light of this, we propose utilizing a separable $L^2$–space on this Lie group, together with the associated Haar measure, as the Hilbert space for our representation. We can employ the Cholesky map to represent the metric components as multiplication operators on this Hilbert space. To represent the conjugate momenta of the spatial metric components, we define a strongly continuous semigroup of shift operators that generate shifts in the positive direction of the metric components. The infinitesimal generators of this semigroup correspond to the momentum operators associated with the metric, and they satisfy the conventional commutation relations. Importantly, the metric operator retains the requirement of positive definiteness in this setting; specifically, $\hat{q}_{ab}s^a s^b$ is positive for all $s \in \mathbb{R}^n$.

Despite satisfying the canonical commutation relations, this representation is not unitarily equivalent to the standard representation of the spatial metric and its conjugate on $L^2(\mathbb{R}^{n(n+1)/2})$. This is not in contradiction with the Stone–von–Neumann theorem, because our canonical commutation relations do not exponentiate to the Weyl algebra, similar to the situation of the radial part of the hydrogen atom. There, the radial component of the atom is also required to be positive, as are some of the degrees of freedom introduced by the Cholesky decomposition. More precisely, these are the diagonal elements of the upper triangular matrices employed for this representation. Similar to the hydrogen atom, the conjugate momenta fail to be (essentially) self–adjoint but the Hamiltonian operator is, as expected, essentially self–adjoint.

Our approach has wide–ranging applicability beyond the specific context of gravitational systems we have considered in this work. Indeed, it is able to obtain non–standard representations of the conventional canonical commutation relations whenever non–trivial configuration spaces need to be incorporated in the quantum framework. Our method is systematic and the complexity of brute–force calculations that are usually employed to arrive at such representations can be bypassed.

Notably, our technique bears similarities to the vielbein approach, wherein local coordinate frames are employed to represent the (spatial) metric tensor (also recognized as "triads" in 3 + 1 gravity). Although these frames facilitate the coupling of spinor fields to gravity, they give rise to supplementary gauge degrees of freedom. Our approach is able to represent such vielbein fields on the same Hilbert space employed for the Cholesky decomposition with a clean separation of physical and gauge degrees of freedom.

Finally, we extend the usual Weyl quantization scheme to our representation of the commutation relations in the most simple and natural way. This allows us to quantize a wide class of functions, including the constraints of gravity. In fact, this Weyl quantization scheme can be used to represent the constraints on a full tensor product Hilbert space composed of the individual Hilbert spaces at each lattice site. We will come to this in an upcoming publication (Lang and

Schander 2023b).

We want to acknowledge that other authors (Isham and Kakas 1984; Klauder 1999) had already obtained a positive–definite quantization of the metric field prior to us. However, their approach differs from ours in that they need to pass to new variables that obey non–canonical commutation relations, whereas our approach works with the standard canonical pair of ADM variables. Moreover, we want to mention the recent publication by Thiemann (2023), who recognized the relevance of the Cholesky decomposition in gravity independently of us, albeit with a different application in mind, i.e. as a gauge fixing method.

With these preliminary remarks, let us introduce the structure of this paper: Section 2 starts by reviewing the quantum case of the hydrogen atom, and as such it provides an intuitive and well–known example for the representation of the gravitational commutation relations that we have in mind. In section 3, we introduce a semigroup of quantum operators that generates shifts in the direction of a generic configuration variable $q$, which is restricted to a configuration space $Q \subset \mathbb{R}^n$, as well as their infinitesimal generators $\hat{p}$. Section 4 applies these results to the case of gravity where the metric degrees of freedom are restricted to $q_{ab}$ being positive definite. We exemplify our findings for the case of $2+1$ and $3+1$ quantum gravity. In section 5, we show that our representation easily extends to include spinor fields. Then, in section 6, we introduce a generalized Weyl quantization formula which applies to our representation of the gravitational quantum degrees of freedom. Finally, section 7 concludes the paper by providing a summary of our findings and an outlook to future investigations.

## 2 Warm up Example: The Hydrogen Atom

To provide a simple example as a warm up for the later calculation, we demonstrate our method in the case of the radial part of the hydrogen atom. Many of the issues we will be facing later, are already contained in this simple example.

In the case of the hydrogen atom, we are working on the Hilbert space $\mathcal{H} = L^2(\mathbb{R}_0^+, r^2 \, dr)$. The operator $\hat{r}$ corresponding to the radial coordinate can directly be represented on the Hilbert space as $(\hat{r}\psi)(r) = r\psi(r)$ and is only allowed to take non–negative values by construction. We are looking for a momentum operator $\hat{p}$ such that the canonical commutation relation

$$[\hat{r}, \hat{p}] = i \tag{2.1}$$

holds. In principle, infinitely many choices are possible. If $\hat{p}$ satisfies equation (2.1), then so does $\hat{p}' = \hat{p} + f(\hat{r})$ for infinitely many choices of $f$.

Our method rests on the idea that the momentum operator should be the generator of translations in the direction of the canonically conjugate configuration variable. So instead of looking for an operator that satisfies equation (2.1) directly, we define an appropriate shift operator and derive the momentum operator as its generator. Not only does this produce an operator that satisfies equation (2.1), but it also seems to pick the physically correct one among the infinitely many possible choices. While there are simpler methods to find the radial momentum operator in the case of the hydrogen atom, our method generalizes well to more complicated and higher–dimensional situations, as we will see later.

Let us first introduce the notion of a contraction semigroup.

**Definition 2.1** (Contraction Semigroup). A family $\{\, T(s) \mid s \geq 0 \,\} \subseteq B(\mathcal{H})$ of bounded operators on $\mathcal{H}$ is called a contraction semigroup if the following conditions hold:
  (i) $T(0) = \mathrm{id}$,
  (ii) $T(s)T(s') = T(s + s')$ for all $s, s' \geq 0$,
  (iii) The map $s \mapsto T(s)\psi$ is continuous for each $\psi \in \mathcal{H}$,
  (iv) $\|T(s)\| \leq 1$ for all $s \geq 0$.

We define a family $\{\, U(s) \mid s \geq 0 \,\} \subseteq B(\mathcal{H})$ of operators that generates shifts in $r$–direction by

$$(U(s)\psi)(r) := \frac{r+s}{r}\psi(r+s). \tag{2.2}$$

The role of the prefactor will become clear shortly. We now check whether this constitutes a contraction semigroup. Items (i) to (iii) in definition 2.1 are true by inspection. We can use integration by substitution to show that $U(s)$ is indeed a contraction,

$$\|U(s)\psi\|^2 = \int_0^\infty \left|\frac{r+s}{r}\psi(r+s)\right|^2 r^2 \, \mathrm{d}r = \int_0^\infty |\psi(r+s)|^2 (r+s)^2 \, \mathrm{d}r$$
$$= \int_s^\infty |\psi(r)|^2 r^2 \, \mathrm{d}r \leq \int_0^\infty |\psi(r)|^2 r^2 \, \mathrm{d}r = \|\psi\|^2. \tag{2.3}$$

We note that unless $\psi(r) = 0$ for $r \leq s$, $U(s)$ is a true contraction and does not preserve the norm. Thus, it is not a unitary operator and its generator is not self–adjoint (cf. Stone's theorem on strongly continuous unitary one–parameter groups). This is not a drawback of our approach, but rather a necessity, as it reproduces the correct quantum mechanical result. The radial momentum operator is, after all and contrary to popular belief, indeed not self–adjoint.

The kernel of $U(s)$ is given by

$$\ker U(s) = \{\, \psi \in \mathcal{H} \mid \psi(r) = 0 \quad \forall r > s \,\}, \tag{2.4}$$

as can be seen from equation (2.3). For these functions, the integral will vanish, yielding a zero norm. Otherwise, the integral will be nonzero. When restricted to $(\ker U(s))^\perp$, equality in equation (2.3) holds, thus making $U(s)$ into a partial isometry. More precisely, we have that $U(s)U(s)^\dagger = \mathrm{id}$ and $U(s)^\dagger U(s) = P_{(\ker U(s))^\perp}$ where $P_{(\ker U(s))^\perp}$ is the projection onto $(\ker U(s))^\perp$. This is ensured by the prefactor in equation (2.2).

We can now move on to calculate the generator of the contraction semigroup $U(s)$.

**Definition 2.2** (Infinitesimal Generator of a Contraction Semigroup). Let $\{\, T(s) \mid s \geq 0 \,\} \subseteq B(\mathcal{H})$ be a contraction semigroup. A closed, densely defined operator $A \colon D(A) \to \mathcal{H}$ is called a generator of $T(s)$ if
  (i) $D(A) = \{\, \psi \in \mathcal{H} \mid \lim_{s \to 0} s^{-1}(T(s)\psi - \psi) \text{ exists} \,\}$,
  (ii) $A\psi = ((\mathrm{d}/\mathrm{d}s)T(s)\psi)_{s=0}$.

The relation between contraction semigroups and their infinitesimal generators is characterized by the follwing version of the theorem of Hille and Yosida (cf. Hille and Phillips 1957).

**Theorem 2.3** (Hille–Yosida)**.** *An operator $A$ is an infinitesimal generator of a contraction semigroup $\{T(s) \mid s \geq 0\} \subseteq B(\mathcal{H})$ if and only if*

(i) $\mathbb{R}^+$ *is contained in the resolvent set of $A$,*

(ii) $\|(A - \lambda)^{-1}\| \leq \lambda^{-1}$ *for all $\lambda > 0$.*

This allows us to compute the infinitesimal generator associated with the contraction semigroup $U(s)$. According to definition 2.2, the generator is given by

$$\mathrm{i}(\hat{p}\psi)(r) = \left(\frac{\mathrm{d}(U(s)\psi)(r)}{\mathrm{d}s}\right)_{s=0} = \frac{\partial \psi}{\partial r}(r) + \frac{1}{r}\psi(r). \tag{2.5}$$

Note that since only positive values of $r$ are allowed, the shift operator cannot be invertible and hence not unitary. In fact, it can easily be checked that the step function $\chi_{[0,s)}$ is in the kernel of $U(s)$. Thus, its generator $\hat{p}$ is only symmetric and not self–adjoint. This can also be seen by computing the deficiency indices of $\hat{p}$, i.e., the dimensions of $\ker(\mathrm{i}\,\mathrm{id} \pm \hat{p}^\dagger)$, (Reed and Simon 1975). The deficiency index associated with $(\mathrm{i}\,\mathrm{id} + \hat{p}^\dagger)$ is simply zero because the solution $\psi_+$ of $(\mathrm{i}\,\mathrm{id} + \hat{p}^\dagger)\psi_+ = 0$ fails to be in the domain of definition of this operator. In fact, we have $\psi_+(r) = ce^r/r$ for some arbitrary $c \in \mathbb{C}$, which is not normalizable. On the other hand, the solution $\psi_-$ of $(\mathrm{i}\,\mathrm{id} - \hat{p}^\dagger)\psi_- = 0$ is $\psi_-(r) = ce^{-r}/r$ which is normalizable. The deficiency index is one. Since the two indices fail to be identical, $\hat{p}$ is not self–adjoint and there are no self–adjoint extensions of $\hat{p}$.

At the same time, it is straightforward to check that $\hat{p}$ together with $\hat{r}$ satisfy the standard commutation relations

$$[\hat{r}, \hat{p}] = \left[r, -\mathrm{i}\left(\frac{\partial}{\partial r} + \frac{1}{r}\right)\right] = \left[r, -\mathrm{i}\frac{\partial}{\partial r}\right] = \mathrm{i}. \tag{2.6}$$

Finally, let us examine the kinetic part of the Hamiltonian. With mass $m = 1$, it reads

$$\hat{H}_{\mathrm{kin}} = \frac{\hat{p}^2}{2} = -\frac{1}{r^2}\frac{\partial}{\partial r}r^2\frac{\partial}{\partial r}. \tag{2.7}$$

By computing $\hat{H}^\dagger$ and the deficiency indices of $\hat{H}$ in the same manner as above, it can easily be checked that $\hat{H}_{\mathrm{kin}}$ is not only symmetric but also admits a self–adjoint extension as one would expect.

## 3   Generalized Shift Operators and Momenta

We now want to generalize the method illustrated in the previous section to higher dimensions and more general configuration spaces. The basic setting will be as follows:

We consider the Hilbert space $\mathcal{H} = L^2(X, \rho(x)\,\mathrm{d}x)$ with $X \subseteq \mathbb{R}^n$. For $Q \subseteq \mathbb{R}^n$, let $q: X \to Q$ be a diffeomorphism. We require that $Q + \mathbb{R}^n_+ \subseteq Q$. Let the shift function $g_s: X \to X$ be given by $g_s(x) = q^{-1}(q(x) + s)$, where $s \in \mathbb{R}^n_+$. Occasionally, we also allow $s \in \mathbb{R}^n$. However, we must then ensure that $q(x) + s \in Q$, which is not always the case for every pair $(x, s)$. Moreover, we can define the multiplication operators $\hat{q}_i : D(\hat{q}_i) \to \mathcal{H}$ by $(\hat{q}_i \psi)(x) = q_i(x) \psi(x)$, where $D(\hat{q}_i) = \{ \psi \in \mathcal{H} \mid \|\hat{q}_i \psi\| < \infty \}$.

Our objective is to define a contraction semigroup $\{ U(s) \in B(\mathcal{H}) \mid s \in \mathbb{R}^n_+ \}$ of partial isometries that generates shifts in $q$ direction, i.e. $U(s)\hat{q}_i U(s)^\dagger = \hat{q}_i + s_i$. (Strictly speaking, only one–parameter semigroups can be contraction semigroups. By abuse of notation, we apply the term to the more general object as well. One can always restrict to a ray in $\mathbb{R}_+$ in order to obtain a true contraction semigroup.)

We are furthermore interested in the infinitesimal generators $\hat{p}_i$ of $U(s)$. Together with the $\hat{q}_i$, they will form a non-standard representation of the canonical commutation relations, just as in the case of the hydrogen atom.

Let us first define the notion of a generalized shift operator.

**Definition 3.1** (Generalized Shift Operator). For $s \in \mathbb{R}^n_+$, the generalized shift operator $U(s)$ in the direction of $q$ is given by the following expression:

$$(U(s)\psi)(x) = \sqrt{\det(J_{g_s}(x)) \frac{\rho(g_s(x))}{\rho(x)}} \psi(g_s(x)). \tag{3.1}$$

**Remark 3.2** (Alternative form of the Generalized Shift Operator). An alternative way to write the operator $U(s)$, involving only the Jacobian of $q$, is given by

$$(U(s)\psi)(x) = \sqrt{\frac{\det J_q(x)}{\det J_q(g_s(x))} \frac{\rho(g_s(x))}{\rho(x)}} \psi(g_s(x)). \tag{3.2}$$

In order to see this, we use the chain rule as well as the inverse function theorem to get

$$J_{g_s}(x) = J_{q^{-1}}(q(x) + s) J_q(x) = J_q(q^{-1}(q(x) + s))^{-1} J_q(x) = J_q(g_s(x))^{-1} J_q(x)$$

and therefore

$$\det(J_{g_s}(x)) = \frac{\det(J_q(x))}{\det(J_q(g_s(x)))}.$$

The crucial properties of $U(s)$, which are the reason we are interested in this operator, are summarized in the following lemma.

**Lemma 3.3** (Properties of the Generalized Shift Operator). *The generalized shift operator $U(s)$ is a partial isometry. The set $\{ U(s) \in B(\mathcal{H}) \mid s \in \mathbb{R}^n_+ \}$ forms a strongly continuous contraction semigroup.*

*Proof.* Let us first compute the norm of $U(s)\psi$,

$$\|U(s)\psi\|^2 = \int_X |(U\psi)(x)|^2 \rho(x)\, dx = \int_X \left| \sqrt{\det(J_{g_s}(x)) \frac{\rho(g_s(x))}{\rho(x)}} \psi(g_s(x)) \right|^2 \rho(x)\, dx$$

$$= \int_X |\psi(g_s(x))|^2 \rho(g_s(x)) |\det(J_{g_s}(x))| \, dx = \int_{g_s(X)} |\psi(x)|^2 \rho(x) \, dx$$

$$\leq \int_X |\psi(x)|^2 \rho(x) \, dx = \|\psi\|^2.$$

If $g_s(X) = X$, we have $\|U(s)\psi\| = \|\psi\|$ for all $\psi \neq 0$ and $U(s)$ is an isometry. Otherwise, the kernel of $U(s)$ and its orthogonal complement are given by

$$\ker U(s) = \{\, \psi \in \mathcal{H} \mid \psi(x) = 0 \quad \forall x \in g_s(X) \,\},$$
$$(\ker U(s))^\perp = \{\, \psi \in \mathcal{H} \mid \psi(x) = 0 \quad \forall x \in X \setminus g_s(X) \,\},$$

and $U(s)$ is a partial isometry because $\|U(s)\psi\| = \|\psi\|$ for $\psi \in (\ker U(s))^\perp$. Since $\|U(s)\psi\| \leq \|\psi\|$, we have $\|U(s)\| \leq 1$, so $U(s)$ is a contraction. Next, we check the semigroup properties. First, we note that $g_0(x) = x$ and we find

$$(U(0)\psi)(x) = \sqrt{\det(J_{g_0}(x)) \frac{\rho(g_0(x))}{\rho(x)}} \psi(g_0(x)) = \sqrt{\det(J_{\mathrm{id}}(x)) \frac{\rho(x)}{\rho(x)}} \psi(x) = \psi(x).$$

Moreover, we can use the composition rule,

$$g_s(g_t(x)) = q^{-1}(q(q^{-1}(q(x) + t)) + s) = q^{-1}(q(x) + t + s) = g_{s+t}(x),$$

to prove the semigroup property

$$(U(s)U(t)\psi)(x) = \sqrt{\det(J_{g_s}(x)) \frac{\rho(g_s(x))}{\rho(x)}} (U(t)\psi)(g_s(x))$$

$$= \sqrt{\det(J_{g_s}(x)) \frac{\rho(g_s(x))}{\rho(x)}} \sqrt{\det(J_{g_t}(g_s(x))) \frac{\rho(g_t(g_s(x)))}{\rho(g_s(x))}} \psi(g_t(g_s(x)))$$

$$= \sqrt{\det(J_{g_t}(g_s(x)) J_{g_s}(x)) \frac{\rho(g_{s+t}(x))}{\rho(x)}} \psi(g_{s+t}(x))$$

$$= \sqrt{\det(J_{g_t \circ g_s}(x)) \frac{\rho(g_{s+t}(x))}{\rho(x)}} \psi(g_{s+t}(x)) = \sqrt{\det(J_{g_{s+t}}(x)) \frac{\rho(g_{s+t}(x))}{\rho(x)}} \psi(g_{s+t}(x))$$

$$= (U(s+t)\psi)(x). \qquad \square$$

The fact that $U(s)$ is a partial isometry implies that $U(s)U(s)^\dagger = \mathrm{id}$ and $U(s)^\dagger U(s) = P_{(\ker U(s))^\perp}$ is a projection onto $(\ker U(s))^\perp$. Even though these relations often suffice in typical calculations, it may nevertheless sometimes be useful to have an explicit expression for $U(s)^\dagger$.

**Lemma 3.4** (Adjoint of the Generalized Shift Operator). *The adjoint of $U(s)$ is given by:*

$$(U(s)^\dagger \psi)(x) = \begin{cases} \sqrt{\det(J_{g_{-s}}(x)) \frac{\rho(g_{-s}(x))}{\rho(x)}} \psi(g_{-s}(x)), & x \in g_s(X), \\ 0, & \textit{otherwise.} \end{cases} \quad (3.3)$$

*Proof.* First, we note that this expression is well–defined although it involves negative shifts of the form $g_{-s}$. Since these only occur in the case $x \in g_s(X)$, it is always possible to apply a backwards shift by at least an amount of $s$.

The following computation verifies that this is indeed the adjoint of $U(s)$. We use the fact that the integrand is 0 on $X \backslash g_s(X)$ and perform a substitution. Furthermore, we note that $g_{-s} = g_s^{-1}$, which allows us to use the inverse function theorem.

$$\begin{aligned}
\langle U(s)^\dagger \phi, \psi \rangle &= \int_{g_s(X)} \sqrt{\det(J_{g_{-s}}(x)) \frac{\rho(g_{-s}(x))}{\rho(x)}} \phi(g_{-s}(x))^* \psi(x) \rho(x) \, dx \\
&= \int_X \sqrt{\det(J_{g_{-s}}(g_s(x))) \frac{\rho(g_{-s}(g_s(x)))}{\rho(g_s(x))}} \phi(g_{-s}(g_s(x)))^* \psi(g_s(x)) \rho(x) \det(J_{g_s}(x)) \, dx \\
&= \int_X \phi(x)^* \sqrt{\det(J_{g_s}(x))^{-1} \frac{\rho(x)}{\rho(g_s(x))}} \psi(g_s(x)) \rho(g_s(x)) \det(J_{g_s}(x)) \, dx \\
&= \int_X \phi(x)^* \sqrt{\det(J_{g_s}(x)) \frac{\rho(g_s(x))}{\rho(x)}} \psi(g_s(x)) \rho(x) \, dx = \langle \phi, U(s) \psi \rangle . \qquad \square
\end{aligned}$$

We are now ready to prove that $U(s)$ generates shifts in $q$ direction. This is analogous to the action of the translation operators in the standard representation of the canonical commutation relations.

**Lemma 3.5.** *$U(s)$ generates shifts in $q_i$ direction*

$$U(s) \hat{q}_i U(s)^\dagger = \hat{q}_i + s_i. \tag{3.4}$$

*Proof.* The proof works by expanding $U(s)$ and then assembling it back together, while factoring out the shifted multiplication operator. We then insert the definition of $g_s$ and find

$$\begin{aligned}
(U(s) \hat{q}_i U(s)^\dagger \psi)(x) &= \sqrt{\det(J_{g_s}(x)) \frac{\rho(g_s(x))}{\rho(x)}} q_i(g_s(x))(U(s)^\dagger \psi)(g_s(x)) \\
&= q_i(g_s(x))(U(s) U(s)^\dagger \psi)(x) = q_i(q^{-1}(q(x) + s)) \psi(x) \\
&= (q_i(x) + s_i) \psi(x) = ((\hat{q}_i + s_i) \psi)(x). \qquad \square
\end{aligned}$$

Next, we shall turn our attention to the generators of the generalized shift operator. These will later play the role of the conjugate momenta to the operators $\hat{q}_i$. The following theorem provides an explicit expression that only depends on the function $q : X \to Q$ and can be directly evaluated.

**Theorem 3.6** (Generators of the Generalized Shift Operator). *The infinitesimal generators of the generalized shift operator are given by $i\hat{p}_i$, where $\hat{p}_i$ reads*

$$\hat{p}_i = -i(J_q(x)^{-1})_{ji} \left( \frac{\partial}{\partial x_j} - \frac{1}{2} \frac{\partial}{\partial x_j} \log\left( \frac{\det(J_q(x))}{\rho(x)} \right) \right). \tag{3.5}$$

*Proof.* As a first step, we compute the derivatives of $g_s$, making use of the inverse function theorem:

$$\left(\frac{\partial (g_s(x))_j}{\partial s_i}\right)_{s=0} = \left(\frac{\partial}{\partial s_i}(q^{-1})_j(q(x)+s)\right)_{s=0} = \left(\frac{\partial (q^{-1})_j(y)}{\partial y_k}\right)_{y=q(x)} \left(\frac{\partial (q_k(x)+s_k)}{\partial s_i}\right)_{s=0}$$

$$= (J_{q^{-1}}(q(x)))_{jk}\delta_{ki} = (J_q(x)^{-1})_{jk}\delta_{ki} = (J_q(x)^{-1})_{ji}$$

Next, we calculate the derivatives of the fractions in the square root of equation (3.2). The first term becomes:

$$\left(\frac{\partial}{\partial s_i}\frac{\det(J_q(x))}{\det(J_q(g_s(x)))}\right)_{s=0} = -\left(\frac{\det(J_q(x))\frac{\partial}{\partial s_i}\det(J_q(g_s(x)))}{\det(J_q(g_s(x)))^2}\right)_{s=0} = -\frac{\left(\frac{\partial}{\partial s_i}\det(J_q(g_s(x)))\right)_{s=0}}{\det(J_q(x))}$$

$$= -\frac{\frac{\partial}{\partial x_j}\det(J_q(x))}{\det(J_q(x))}\left(\frac{\partial (g_s(x))_j}{\partial s_i}\right)_{s=0} = -\frac{\partial \log(\det(J_q(x)))}{\partial x_j}(J_q(x)^{-1})_{ji}.$$

Similarly, the second term can be computed as follows:

$$\left(\frac{\partial}{\partial s_i}\frac{\rho(g_s(x))}{\rho(x)}\right)_{s=0} = \frac{1}{\rho(x)}\frac{\partial \rho(x)}{\partial x_j}\left(\frac{\partial (g_s(x))_j}{\partial s_i}\right)_{s=0} = \frac{\partial \log(\rho(x))}{\partial x_j}(J_q(x)^{-1})_{ji}.$$

We can now use these derivatives in order to compute the generators of $U(s)$. Note that, in order to apply the Hille–Yosida theorem to prove the existence of a generator, we strictly need a contraction semigroup with only one parameter. Our group is parametrized by $\mathbb{R}^n_+$, but we can, without loss of generality, restrict it to a ray and compute the generator as follows:

$$i\hat{p}_i\psi = \left(\frac{d}{dt}U(te_i)\psi\right)_{t=0} = \left(\frac{\partial}{\partial s_i}\prod_{j=1}^n U(s_je_j)\psi\right)_{s=0} = \left(\frac{\partial}{\partial s_i}U\left(\sum_{j=1}^n s_je_j\right)\psi\right)_{s=0} = \left(\frac{\partial}{\partial s_i}U(s)\psi\right)_{s=0}.$$

Here, $(e_i)_i$ denotes the canonical basis in $\mathbb{R}^n$. We can now perform the following calculation (since $g_0(x) = x$, we can directly evaluate intermediate expressions at $s=0$):

$$i(\hat{p}_i\psi)(x) = \left(\frac{\partial}{\partial s_i}(U(s)\psi)(x)\right)_{s=0} = \left(\frac{\partial}{\partial s_i}\sqrt{\frac{\det J_q(x)}{\det J_q(g_s(x))}\frac{\rho(g_s(x))}{\rho(x)}}\psi(g_s(x))\right)_{s=0}$$

$$= \frac{1}{2}(J_q(x)^{-1})_{ji}\left(\frac{\partial \log(\rho(x))}{\partial x_j} - \frac{\partial \log(\det(J_q(x)))}{\partial x_j}\right)\psi(x) + (J_q(x)^{-1})_{ji}\frac{\partial}{\partial x_j}\psi(x)$$

$$= -\frac{1}{2}(J_q(x)^{-1})_{ji}\left(\frac{\partial}{\partial x_j}\log\left(\frac{\det(J_q(x))}{\rho(x)}\right)\right)\psi(x) + (J_q(x)^{-1})_{ji}\frac{\partial}{\partial x_j}\psi(x)$$

$$= (J_q(x)^{-1})_{ji}\left(\frac{\partial}{\partial x_j}\psi(x) - \frac{1}{2}\left(\frac{\partial}{\partial x_j}\log\left(\frac{\det(J_q(x))}{\rho(x)}\right)\right)\psi(x)\right).$$

From this, equation (3.5) follows. □

After having obtained the generators of the generalized shift operator, we are now interested in its properties. One interesting feature is that they are symmetric. Even though they are not necessarily self–adjoint, this often allows for the self–adjoint quantization of some functions of the corresponding classical variables.

**Lemma 3.7** (Symmetry of the Generators). *The generators of the generalized shift operator are symmetric. If* $\ker U(s) = 0$, *they are in fact self–adjoint.*

*Proof.* Let us define

$$\bar{D}(\hat{p}_i) = D(\hat{p}_i) \cap \left\{ \psi \in \mathcal{H} \mid \exists \varepsilon > 0 : \psi(X \setminus g_{\varepsilon e_i}(X)) = 0 \right\}$$

and assume that $\psi \in \bar{D}(\hat{p}_i)$. Evidently, $\bar{D}(\hat{p}_i)$ is dense in $D(\hat{p}_i)$. We use the fact that $U(s)$ is a partial isometry, i.e. $\|U(s)\psi\| = \|\psi\|$ for $\psi \in (\ker U(s))^\perp$. Since $\psi \in \bar{D}(\hat{p}_i)$, there is some $t_0$ such that $\psi \in (\ker U(te_i))^\perp$ for all $t < t_0$. Hence, the following equality holds:

$$\left(\frac{\mathrm{d}}{\mathrm{d}t}\|U(te_i)\psi\|^2\right)_{t=0} = \left(\frac{\mathrm{d}}{\mathrm{d}t}\langle U(te_i)\psi, U(te_i)\psi\rangle\right)_{t=0} = \left\langle \left(\frac{\mathrm{d}}{\mathrm{d}t}U(te_i)\psi\right)_{t=0}, \psi\right\rangle + \left\langle \psi, \left(\frac{\mathrm{d}}{\mathrm{d}t}U(te_i)\psi\right)_{t=0}\right\rangle$$
$$= \langle \mathrm{i}\hat{p}_i\psi, \psi\rangle + \langle \psi, \mathrm{i}\hat{p}_i\psi\rangle = \left(\frac{\partial}{\partial s_i}\|\psi\|^2\right)_{s=0} = 0$$

From this, the symmetry of $\hat{p}_i$ on $\bar{D}(\hat{p}_i)$ follows by polarization:

$$\langle \hat{p}_i\phi, \psi\rangle = \langle \phi, \hat{p}_i\psi\rangle$$

Since $\bar{D}(\hat{p}_i)$ is dense in $D(\hat{p}_i)$ and $\hat{p}_i$ is closed, this also holds on $D(\hat{p}_i)$. For the case $\ker U(s) = 0$, we note that $(\ker U(s))^\perp = \mathcal{H}$, so $U(s)^\dagger U(s) = \mathrm{id}$ and $U(s)$ is in fact unitary. The the adjoint in lemma 3.4 is then the inverse of $U(s)$. Thus, the Hille–Yosida theorem used in the proof of theorem 3.6 specializes to Stone's theorem on unitary one-parameter groups. Therefore, the $\hat{p}_i$ are self–adjoint in this situation. □

The next theorem can be considered the central result of this section. It shows that our method of generalized shift operators allows one to compute canonically conjugate momenta to a very wide class of configuration variables in any dimension. Although we have applications in gravity in mind, the method is potentially useful for many other fields as well.

**Theorem 3.8** (Canonical Commutation Relations). *The generators of the generalized shift operator satisfy the canonical commutation relations:*

$$[\hat{q}_i, \hat{p}_j] = \mathrm{i}\delta_{ij}, \quad [\hat{q}_i, \hat{q}_j] = [\hat{p}_i, \hat{p}_j] = 0 \tag{3.6}$$

*Proof.* The commutator between the $\hat{q}_i$ components is trivial, since the $\hat{q}_i$ act as multiplication operators. For the commutator between the $\hat{p}_i$ components, we note that:

$$[\hat{p}_i, \hat{p}_j] = -\left(\frac{\partial^2}{\partial s_i\, \partial t_j}(U(s)U(t) - U(t)U(s))\right)_{s,t=0} = -\left(\frac{\partial^2}{\partial s_i\, \partial t_j}(U(s+t) - U(s+t))\right)_{s,t=0} = 0$$

The commutator between $\hat{q}_i$ and $\hat{p}_j$ follows from

$$\delta_{ij} = \left(\frac{\partial}{\partial s_j}(\hat{q}_i + s_i)\right)_{s=0} = \left(\frac{\partial}{\partial s_j}U(s)\hat{q}_i U(s)^\dagger\right)_{s=0} = \left(\frac{\partial}{\partial s_j}U(s)\right)_{s=0}\hat{q}_i + \hat{q}_i\left(\frac{\partial}{\partial s_j}U(s)^\dagger\right)_{s=0}$$
$$= (\mathrm{i}\hat{p}_j)\hat{q}_i + \hat{q}_i(-\mathrm{i}\hat{p}_j^\dagger)$$

where lemma 3.5 was invoked. Since the restriction of $\hat{p}_j^\dagger$ to $D(\hat{p}_j)$ is just $\hat{p}_j$, we can conclude that $[\hat{q}_i, \hat{p}_j] = \mathrm{i}\delta_{ij}$. □

## 4  Positive Definite Matrices and Conjugate Momenta

In this section, we leverage the previous findings to establish a Hilbert space and a non–standard representation of the canonical commutation relations that are tailored to the requirements of canonical quantum gravity. Usually, one defines the canonical variables on the classical phase space as fields $q_{ab}(x)$ and $p^{cd}(y)$ that exhibit symmetry in their indices,

$$\{q_{ab}(x), p^{cd}(y)\} = \delta^a{}_{(a}\delta^d{}_{b)}\delta(x-y). \tag{4.1}$$

Although in principle, Einstein's field equations work quite well without any further qualifications, from a physical perspective, one would like the spatial metric to be Riemannian, which amounts to the additional requirement that $q_{ab}(x)$ is positive definite. Enforcing this requirement in the context of quantum gravity is not a straightforward task.

Putting aside the potential issues related to the quantization of field theories, let us consider the situation on a spatial lattice. In such a case, it is desirable to have a positive definite metric on each lattice site. Quantum mechanically, the Hilbert space of the lattice theory can be defined as a tensor product of Hilbert spaces on each lattice site. Each site Hilbert space should host only quantum states that have support on positive definite symmetric matrices, i.e., there should be a representation $\hat{q}_{ab}$, $\hat{p}^{ab}$ of the canonical variables satisfying the canonical commutation relations with the additional requirement that $\hat{q}_{ab}s^a s^b$ is a positive operator for every $s \in \mathbb{R}^f$, where $f = n(n+1)/2$ is the number of independent degrees of freedom of a symmetric matrix.

In the conventional representation of the canonical commutation relations, where $\mathcal{H} = L^2(\mathbb{R}^f)$ and $\hat{q}_{ab}$ and $\hat{p}^{ab}$ act as multiplication operators and partial derivatives with respect to $q_{ab}$, this feature is absent. In this representation, the elements $\hat{q}_{ab}$ are independent and can assume values across $\mathbb{R}^f$. As a result, the enforcement of positive definiteness is not guaranteed.

### 4.1  Cholesky Decomposition and the Lie Group of Upper Triangular Matrices

Rather than employing $\mathbb{R}^f$ as the configuration space, we prefer to adopt the space of positive definite symmetric matrices, which represents a subset of $\mathbb{R}^{n^2}$, defined by intricate equations. To achieve quantization in this space, it is first necessary to parametrize it. Among the possible parametrizations, the Cholesky decomposition is a widely used approach, (cf. Golub and Van Loan 2013).

**Theorem 4.1** (Cholesky Decomposition)*. Every positive definite matrix A can be decomposed into the product $A = U^\mathsf{T} U$, where U is an upper triangular matrix with positive diagonal elements. This decomposition is unique.*

*Proof.* Since $A$ is symmetric and positive definite, there exists a unique positive definite square root $B$ such that $A = B^\mathsf{T} B$. Let $B = OU$ be its QR–decomposition, where $O$ is an orthogonal matrix and $U$ is upper triangular. Potential negative signs on the diagonal of $U$ can be shifted onto $O$. Thus, $U$ satisfies the conclusions of the theorem. It is now easy to see that

$$A = B^\mathsf{T} B = U^\mathsf{T} O^\mathsf{T} O U = U^\mathsf{T} U, \tag{4.2}$$

where $O^\mathsf{T} O = \mathrm{id}$ has been used. Uniqueness follows from the uniqueness of the square root and the fact that there is only one way to shift potential minus signs onto $O$. □

We can thus take the space of upper triangular matrices with positive diagonal elements as our configuration space. But what would be an appropriate measure? On the one hand, one could of course choose the Lebesgue measure, but this does not do justice to the fact that the space of symmetric, positive definite matrices really has the structure of a complicated manifold and the Choleksy decomposition gives rise to just one way of parametrizing it. A more invariant way to specify a measure is to realize that the set of upper diagonal matrices with positive diagonal elements carries the structure of a Lie group. A natural choice of measure is therefore provided by the Haar measure.

**Definition 4.2** (Group of Upper Triangular Matrices with Positive Diagonal Elements)**.** The group of upper triangular real $n \times n$ matrices with positive diagonal elements will be denoted by

$$\mathrm{UT}_+(n, \mathbb{R}) := \{\, u \in GL(n, \mathbb{R}) \mid \forall i > j \colon u_{ij} = 0 \wedge \forall i \colon u_{ii} > 0 \,\}. \tag{4.3}$$

It is easy to check that this indeed defines a group. Since this group is specified by a set of polynomial equations, it forms a Lie group by the closed subgroup theorem, while the inequalities just restrict the Lie group to the connected component of the identity.

We can now calculate the Haar measure on $\mathrm{UT}_+(n, \mathbb{R})$. On this group, the left Haar measure is not equal to the right Haar measure, so we are faced with a choice. We will use the left Haar masure here because it plays well together with the Cholesky decomposition, as we will see shortly. If we had instead used lower triangular matrices $L$ and the alternative Cholesky decomposition $A = LL^\mathsf{T}$, then the right Haar measure would fit accordingly.

**Lemma 4.3** (Left Haar Measure on $\mathrm{UT}_+(n, \mathbb{R})$)*. Up to an arbitrary constant, the left Haar measure* $\mathrm{d}\mu = \rho(u)\,\mathrm{d}u$ *on* $\mathrm{UT}_+(n, \mathbb{R})$*, where* $\mathrm{d}u = \prod_{i \leq j} \mathrm{d}u_{ij}$ *and* $(u_{ij})_{i \leq j}$ *is the coordinate chart of matrix elements, is given by the following density:*

$$\rho(u) = \left( \prod_{k=1}^{n} (u_{kk})^{n-k+1} \right)^{-1} \tag{4.4}$$

*Proof.* By the left invariance of the Haar measure and integration by substitution, we find that

$$\int_X \rho(u)\,du = \int_{l_g(X)} \rho(u)\,du = \int_X \rho(l_g(u))\,|\det(J_{l_g}(u))|\,du.$$

Here, $l_g$ denotes the left translation $l_g(u) = gu$. Since this must hold for all $X \subseteq \mathrm{UT}_+(n,\mathbb{R})$, it follows that the integrands must be equal. Evaluating them at $u = e$ yields

$$\rho(g) = |\det(J_{l_g}(e))|^{-1}.$$

We thus need to compute the Jacobian of the left translation. For $\mathrm{UT}_+(n,\mathbb{R})$, it can be specified in terms of its matrix elements by the following expression:

$$(l_g(u))_{ij} = \begin{cases} \sum_{i \leq k \leq j} g_{ik} u_{kj}, & \forall i \leq j, \\ 0, & \text{otherwise.} \end{cases}$$

For the rest of the proof, we generally assume $i \leq j$ and $k \leq l$. It is then easy to calculate the partial derivatives

$$\left(\frac{\partial (l_g(u))_{ij}}{\partial u_{kl}}\right)_{u=e} = \sum_{i \leq m \leq j} g_{im} \delta_{mk} \delta_{jl} = \begin{cases} g_{ik}, & i \leq k \wedge k \leq j \wedge j = l, \\ 0, & \text{otherwise.} \end{cases}$$

In order to arrange the derivatives $\partial (l_g(u))_{ij}/\partial u_{kl}$ into a Jacobian matrix, an ordering on the indices must be chosen. We chose the lexicographical order, which amounts to

$$(i,j) \leq (k,l) \quad \text{if and only if} \quad i < k \vee (i = k \wedge j \leq l).$$

We now show that the Jacobian is upper triangular with respect to this ordering. The Jacobian determinant thus becomes the product of the diagonal elements. This amounts to showing that $(k,l) < (i,j)$ implies $\neg(i \leq k \wedge k \leq j \wedge j = l)$, i.e., the entries below the diagonal are not vanishing. Equivalently, we may prove the contrapositive statement

$$i \leq k \wedge k \leq j \wedge j = l \implies i < k \vee (i = k \wedge j \leq l),$$

where $(k,l) < (i,j) \equiv \neg((i,j) \leq (k,l))$ has been used, the double negation has been eliminated and the definition of the lexicographical order has been substitituted. It is now easy to see that $i \leq k$ and $j = l$ taken together make the disjunction on the right hand side true. Therefore, the Jacobian is upper triangular and the Jacobian determinant is given by

$$\det(J_{l_g}(e)) = \prod_{i \leq j} \left(\frac{\partial(l_g(u))_{ij}}{\partial u_{ij}}\right)_{u=e} = \prod_{i \leq j} g_{ii} = \prod_{k=1}^n (g_{kk})^{n-k+1}.$$

From this and the fact that the diagonal elements are positive, the lemma follows. □

This concludes our discussion of the Hilbert space, which shall be understood to be $\mathcal{H} = L^2(\mathrm{UT}_+(n,\mathbb{R}), \rho(u)\,du)$ in the remainder of the paper. We can now proceed to provide explicit formulas for the generalized shift operator (equation (3.2)) and generalized momentum operators (equation (3.5)).

*4.2 Generalized Momentum Operators for the Cholesky Decomposition*

As explained earlier, the spatial metric is a positive definite, symmetric 2–form and the Cholesky decomposition (theorem 4.1) may be used to express it in terms of an upper triangular matrix according to the function $q(u) = u^\intercal u$. In order to compute the generalized shift and momentum operators, the missing ingredient is the Jacobian of this function and its determinant. The following theorem provides these ingredients in the general $n$–dimensional case.

**Lemma 4.4** (Jacobi determinant of Cholesky decomposition)**.** *The Jacobian determinant of the Cholesky decomposition $q(u) = u^\intercal u$, where $u$ is an $n$-dimensional upper triangular matrix with positive diagonal elements, is given by*

$$\det(J_q(u)) = 2^n \prod_{k=1}^{n}(u_{kk})^{n-k+1}$$

*Proof.* The matrix elements of $q(u)$ are given by

$$q_{ij}(u) = \sum_{m=1}^{i} u_{mi} u_{mj}. \tag{4.5}$$

In the following, we will again assume that $i \leq j$ and $k \leq l$. The elements of the Jacobian are then given by

$$\frac{\partial q_{ij}(u)}{\partial u_{kl}} = \sum_{i}(\delta_{mk}\delta_{il}u_{mj} + u_{mi}\delta_{mk}\delta_{jl}) = \begin{cases} \delta_{il}u_{kj} + \delta_{jl}u_{ki} & k \leq i \\ 0 & \text{otherwise} \end{cases}$$

$$= \begin{cases} 2u_{kl} & k \leq i \wedge i = j \wedge (i = l \vee j = l) \\ u_{kl} & k \leq i \wedge i \neq j \wedge (i = l \vee j = l) \\ 0 & k > i \vee (i \neq l \wedge j \neq l) \end{cases}. \tag{4.6}$$

In order to compute its determinant, we show that the Jacobian is lower diagonal. The determinant is therefore again given by the product of the diagonal elements. Again, the derivatives in equation (4.6) will be arranged in lexicographical ordering. We thus have to show that

$$(i,j) < (k,l) \implies k > i \vee (i \neq l \wedge j \neq l),$$

which amounts to showing that the elements above the diagonal are equal to 0. The equivalent contrapositive statement is given by

$$k \leq i \wedge (i = l \vee j = l) \implies k < i \vee (k = i \wedge l \leq j)$$

where $\neg((i,j) < (k,l)) \equiv (k,l) \leq (i,j)$ was used and the definition of the lexicographical order was inserted. This is true, because in the case $k < i$, we satisfy the first part of the disjunction and in the case $k = i$, $l \leq j$ is true because either $i = l$, so $i \leq j$, which is true, or $j = l$, which amounts to $j \leq j$, which is also true. Consequently, the Jacobian is indeed diagonal and we can now compute its determinant:

$$\det(J_q(u)) = \prod_{i \leq j} \frac{\partial q_{ij}(u)}{\partial u_{ij}} = \Big(\prod_{i=j} u_{ii}\Big)\Big(\prod_{i<j} u_{ii}\Big)$$
$$= \Big(\prod_{k=1}^n 2u_{kk}\Big)\Big(\prod_{k=1}^n (u_{kk})^{n-k}\Big) = 2^n \prod_{k=1}^n (u_{kk})^{n-k+1}. \qquad \square$$

We are now ready to write down the generalized shift operator and the generalized momenta conjugate to the multiplication operators $(\hat{q}_{ij}\psi)(u) = q_{ij}(u)\psi(u)$.

**Theorem 4.5.** *The generalized conjugate momenta to $\hat{q}_{ij}$ on $\mathcal{H} = L^2(\mathrm{UT}_+(n, \mathbb{R}), \rho(u)\,\mathrm{d}u)$, where $\rho(u)$ is given by lemma 4.3, read*

$$\hat{p}_{ij} = -\mathrm{i} \sum_{k<l} (J_q(u)^{-1})_{klij}\Big(\frac{\partial}{\partial u_{kl}} - \frac{\partial \log(\det(J_q(u)))}{\partial u_{kl}}\Big). \qquad (4.7)$$

*Proof.* By inspection of lemma 4.4 and lemma 4.3, we note that

$$\rho(u) = 2^{-n} \det(J_q(u))^{-1}.$$

However, one requirement from section 3 was that $Q + \mathbb{R}_+^n \subseteq Q$, so we can't directly use $q(u) = u^\intercal u$ in equations (3.2) and (3.5). If $i \neq j$, then shifting the elements $q_{ij}$ too far into the positive direction will eventually make a positive definite matrix indefinite. However, we may always add matrices of the form $vv^\intercal$ to a positive definite matrix without making it indefinite. By taking the canonical vectors $e_i$ and their sums $e_i + e_j$ for $i \neq j$ as vectors $v$, we use a constant basis change matrix $A$ in order to pass to a basis where the criterion $Q + \mathbb{R}_+^n \subseteq Q$ can always be ensured:

$$g_{As}^q(u) = q^{-1}(q(u) + As) = q^{-1}(A(A^{-1}q(u) + s)) = (A^{-1}q)^{-1}((A^{-1}q)(u) + s) = g_s^{A^{-1}q}(u).$$

We note that $J_{A^{-1}q}(u) = A^{-1}J_q(u)$ and $\det(J_{A^{-1}q}(u)) = \det(A^{-1})\det(J_q(u))$. Here, $(J_q(u))_{klij}$ is given by equation (4.6). Thus, we have

$$(J_{A^{-1}q}(u)^{-1})_{klij} = \sum_{m<n} A_{mnij}(J_q(u)^{-1})_{klmn}.$$

These results can be inserted into equation (3.5) to attain the canonical conjugate momenta to $(A^{-1}\hat{q})_{ij}$:

$$\hat{p}_{ij}^{A^{-1}q} = -\mathrm{i} \sum_{k<l} \sum_{m<n} A_{mnij}(J_q(u)^{-1})_{klmn}\Big(\frac{\partial}{\partial u_{kl}} - \frac{\partial \log(\det(J_q(u)))}{\partial u_{kl}}\Big). \qquad (4.8)$$

It then follows immediately the conjugate momenta to $\hat{q}_{ij}$ are given by equation (4.7), because $A^\intercal$ and $A^{-1}$ cancel in the canonical commutation relations. $\qquad \square$

**Remark 4.6.** We note that in the special case, where $\mathrm{UT}_+(n, \mathbb{R})$ is equipped with the left Haar measure, the measure harmonizes well with the upper triangular Cholesky decomposition. A similar simplification happens for the combination of the right Haar measure and the lower triangular Cholesky decomposition. In that sense, a Haar measure seems to be the most natural choice on $\mathcal{H}$, not only from the perspective of Lie group theory, but also from the perspective of the theory of generalized shift operators.

*4.3 Application to Two and Three Spatial Dimensions*

The following discussion serves to showcase our formalism in the two– and three–dimensional case. While the latter case corresponds to the physically interesting situation, the former case is relevant only in a toy model.

*4.3.1 Two Dimensions*

In two dimensions, the metric $q_{ij}$ is given by a $2 \times 2$ symmetric, positive definite matrix. We can use the Cholesky decomposition $q(u) = u^\intercal u$ to express its matrix elements in terms of the elements of the upper triangular matrix $u$ as follows (cf. equation (4.5)):

$$q_{11} = u_{11}^2, \tag{4.9}$$

$$q_{12} = u_{11}u_{12}, \tag{4.10}$$

$$q_{22} = u_{12}^2 + u_{22}^2. \tag{4.11}$$

It is then straightforward to compute the Jacobian of the map $q(u)$ as well as its inverse. The Jacobian reads

$$J_q(u) = \begin{pmatrix} 2u_{11} & 0 & 0 \\ u_{12} & u_{11} & 0 \\ 0 & 2u_{12} & 2u_{22} \end{pmatrix} \tag{4.12}$$

as can also be seen from equation (4.6). Its determinant is therefore given by

$$\det(J_q(u)) = 4u_{11}^2 u_{22} \tag{4.13}$$

and its non–zero partial derivatives read

$$\frac{\partial \det(J_q(u))}{\partial u_{11}} = 8u_{11}u_{22}, \quad \frac{\partial \det(J_q(u))}{\partial u_{22}} = 4u_{11}^2. \tag{4.14}$$

The inverse of the Jacobian can easily be computed to be

$$J_q(u)^{-1} = \frac{1}{2u_{11}^2 u_{22}} \begin{pmatrix} u_{11}u_{22} & 0 & 0 \\ -u_{12}u_{22} & 2u_{11}u_{22} & 0 \\ u_{12}^2 & -2u_{11}u_{12} & u_{11}^2 \end{pmatrix}. \tag{4.15}$$

By inserting these results into equation (4.7), we obtain the generalized conjugate momenta in the two dimensional case as follows:

$$i\hat{p}_{11} = \frac{1}{2u_{11}}\frac{\partial}{\partial u_{11}} - \frac{u_{12}}{2u_{11}^2}\frac{\partial}{\partial u_{12}} + \frac{u_{12}^2}{2u_{11}^2 u_{22}}\frac{\partial}{\partial u_{22}} - \frac{2u_{22}^2 + u_{12}^2}{2u_{11}^2 u_{22}^2}, \tag{4.16}$$

$$i\hat{p}_{12} = \frac{1}{u_{11}}\frac{\partial}{\partial u_{12}} - \frac{u_{12}}{u_{11}u_{22}}\frac{\partial}{\partial u_{22}} + \frac{u_{12}}{u_{11}u_{22}^2}, \tag{4.17}$$

$$i\hat{p}_{22} = \frac{1}{2u_{22}}\frac{\partial}{\partial u_{22}} - \frac{1}{2u_{22}^2}. \tag{4.18}$$

*4.4 Three Dimensions*

We now repeat the above calculation in three dimensions. Again, we express the metric in terms its Cholesky decomposition. The matrix elements now read

$$q_{11} = u_{11}^2, \qquad (4.19)$$
$$q_{12} = u_{11}u_{12}, \qquad (4.20)$$
$$q_{13} = u_{11}u_{13}, \qquad (4.21)$$
$$q_{22} = u_{12}^2 + u_{22}^2, \qquad (4.22)$$
$$q_{23} = u_{12}u_{13} + u_{22}u_{23}, \qquad (4.23)$$
$$q_{33} = u_{13}^2 + u_{23}^2 + u_{33}^2. \qquad (4.24)$$

This again allows us to compute the Jacobian

$$J_q(u) = \begin{pmatrix} 2u_{11} & 0 & 0 & 0 & 0 & 0 \\ u_{12} & u_{11} & 0 & 0 & 0 & 0 \\ u_{13} & 0 & u_{11} & 0 & 0 & 0 \\ 0 & 2u_{12} & 0 & 2u_{22} & 0 & 0 \\ 0 & u_{13} & u_{12} & u_{23} & u_{22} & 0 \\ 0 & 0 & 2u_{13} & 0 & 2u_{23} & 2u_{33} \end{pmatrix} \qquad (4.25)$$

and its determinant

$$\det(J_q(u)) = 8u_{11}^3 u_{22}^2 u_{33}. \qquad (4.26)$$

The non–zero partial derivatives are now given by

$$\frac{\partial \det(J_q(u))}{\partial u_{11}} = 24u_{1,1}^2 u_{2,2}^2 u_{3,3}, \quad \frac{\partial \det(J_q(u))}{\partial u_{22}} = 16u_{1,1}^3 u_{2,2} u_{3,3}, \quad \frac{\partial \det(J_q(u))}{\partial u_{33}} = 8u_{1,1}^3 u_{2,2}^2 \quad (4.27)$$

and the inverse of the Jacobian reads

$$J_q(u)^{-1} = \begin{pmatrix} \frac{1}{2u_{11}} & 0 & 0 & 0 & 0 & 0 \\ -\frac{u_{12}}{2u_{11}^2} & \frac{1}{u_{11}} & 0 & 0 & 0 & 0 \\ -\frac{u_{13}}{2u_{11}^2} & 0 & \frac{1}{u_{11}} & 0 & 0 & 0 \\ \frac{u_{12}^2}{2u_{11}^2 u_{22}} & -\frac{u_{12}}{u_{11} u_{22}} & 0 & \frac{1}{2u_{22}} & 0 & 0 \\ \frac{u_{12}(2u_{13}u_{22}-u_{12}u_{23})}{2u_{11}^2 u_{22}^2} & \frac{u_{12}u_{23}-u_{13}u_{22}}{u_{11} u_{22}^2} & -\frac{u_{12}}{u_{11} u_{22}} & -\frac{u_{23}}{2u_{22}^2} & \frac{1}{u_{22}} & 0 \\ \frac{(u_{13}u_{22}-u_{12}u_{23})^2}{2u_{11}^2 u_{22}^2 u_{33}} & \frac{u_{23}(u_{13}u_{22}-u_{12}u_{23})}{u_{11} u_{22}^2 u_{33}} & \frac{u_{12}u_{23}-u_{13}u_{22}}{u_{11} u_{22} u_{33}} & \frac{u_{23}^2}{2u_{22}^2 u_{33}} & -\frac{u_{23}}{u_{22} u_{33}} & \frac{1}{2u_{33}} \end{pmatrix}.$$

We are now ready to put together the expressions for the canonical conjugate momenta in the three dimensional case. This yields the following results:

$$i\hat{p}_{11} = \frac{1}{2u_{11}} \frac{\partial}{\partial u_{11}} - \frac{u_{12}}{2u_{11}^2} \frac{\partial}{\partial u_{12}} - \frac{u_{13}}{2u_{11}^2} \frac{\partial}{\partial u_{13}} + \frac{u_{12}^2}{2u_{11}^2 u_{22}} \frac{\partial}{\partial u_{22}} - \frac{u_{12}(u_{12}u_{23} - 2u_{13}u_{22})}{2u_{11}^2 u_{22}^2} \frac{\partial}{\partial u_{23}}$$

$$+ \frac{(u_{13}u_{22} - u_{12}u_{23})^2}{2u_{11}^2 u_{22}^2 u_{33}} \frac{\partial}{\partial u_{33}} - \frac{1}{2u_{11}^2}\left(3 + \frac{u_{12}^2}{u_{22}^2} + \frac{(u_{13}u_{22} - u_{12}u_{23})^2}{u_{22}^2 u_{33}^2}\right), \tag{4.28}$$

$$i\hat{p}_{12} = \frac{1}{u_{11}} \frac{\partial}{\partial u_{12}} - \frac{u_{12}}{u_{11} u_{22}} \frac{\partial}{\partial u_{22}} - \frac{(u_{13}u_{22} - u_{12}u_{23})}{u_{11} u_{22}^2} \frac{\partial}{\partial u_{23}}$$

$$- \frac{u_{23}(u_{12}u_{23} - u_{13}u_{22})}{u_{11} u_{22}^2 u_{33}} \frac{\partial}{\partial u_{33}} - \frac{1}{u_{11} u_{22}^2}\left(\frac{u_{23}(u_{13}u_{22} - u_{12}u_{23}) - 2u_{12}}{u_{33}}\right), \tag{4.29}$$

$$i\hat{p}_{13} = \frac{1}{u_{11}} \frac{\partial}{\partial u_{13}} - \frac{u_{12}}{u_{11} u_{22}} \frac{\partial}{\partial u_{23}} - \frac{u_{13}u_{22} - u_{12}u_{23}}{u_{11} u_{22} u_{33}} \frac{\partial}{\partial u_{33}} - \frac{u_{12}u_{23} - u_{13}u_{22}}{u_{11} u_{22} u_{33}^3}, \tag{4.30}$$

$$i\hat{p}_{22} = \frac{1}{2u_{22}} \frac{\partial}{\partial u_{22}} - \frac{u_{23}}{2u_{22}^2} \frac{\partial}{\partial u_{23}} + \frac{u_{23}^2}{2u_{22}^2 u_{33}} \frac{\partial}{\partial u_{33}} - \frac{1}{u_{22}^2}\left(1 + \frac{u_{23}^2}{2u_{33}^2}\right), \tag{4.31}$$

$$i\hat{p}_{23} = \frac{1}{u_{22}} \frac{\partial}{\partial u_{23}} - \frac{u_{23}}{u_{22} u_{33}} \frac{\partial}{\partial u_{33}} + \frac{u_{23}}{u_{22} u_{33}^2}, \tag{4.32}$$

$$i\hat{p}_{33} = \frac{1}{2u_{33}} \frac{\partial}{\partial u_{33}} - \frac{1}{2u_{33}^2}. \tag{4.33}$$

## 5 Vielbein Representation

In the previous section, we employed the Cholesky decomposition $q = u^\intercal u$, where $u \in \mathrm{UT}_+(n, \mathbb{R})$ is an upper triangular matrix with positive diagonal elements, in order to represent the spatial metric $q$. However, in the presence of fermions, it is often necessary to work in the vielbein formalism, where $q$ is represented in terms of vielbein fields $e^i{}_a$, such that

$$q_{ab} = \delta_{ij} e^i{}_a e^j{}_b \tag{5.1}$$

or in matrix notation

$$q = e^\intercal e \tag{5.2}$$

where $e \in \mathrm{GL}(n, \mathbb{R})$ is an invertible matrix whose columns are given by the vielbein basis vectors. The metric $q$ is positive definite in this representation as well. However, given an orthogonal matrix $o$, the vielbein $e' = oe$ corresponds to the same metric, since

$$q = e^\intercal e = e^\intercal o^\intercal o e = e'^\intercal e'. \tag{5.3}$$

Thus, this representation encompasses additional gauge degrees of freedom. As explained in the proof of theorem 4.1, one may employ the QR–decomposition in order to write $e$ as a product of an orthogonal matrix $o \in \mathrm{O}(n)$ and an upper triangular matrix $u \in \mathrm{UT}_+(n, \mathbb{R})$ with positive diagonal elements, thus fixing a particular choice of gauge.

Since we would like to be able to quantize fermions as well, we need a representation not only of the metric variables $q_{ab}$ and their conjuage momenta $p^{ab}$, but also of the vielbein variables $e^i_a$. Since they encompass gauge degrees of freedom, we need to enlarge our Hilbert space in order to accomodate for the additional degrees of freedom. One possible choice to do so is provided by the following definition:

**Definition 5.1** (Hilbert Space for the Vielbein Representation)**.** The Hilbert space for the vielbein representation is given by

$$\mathcal{H} = L^2(\mathrm{UT}_+(n, \mathbb{R}), \rho(u)\,\mathrm{d}u) \otimes L^2(\mathrm{O}(n), \mathrm{d}\mu), \tag{5.4}$$

where $\mathrm{d}\mu$ is the Haar measure on $\mathrm{O}(n)$. The vielbein variables are represented as multiplication operators $\hat{e}^i{}_a$ as follows:

$$(\hat{e}^i{}_a \psi)(u, o) = \sum_j o^i{}_j u_{ja} \psi(u, o). \tag{5.5}$$

**Remark 5.2.** The metric is again represented by $(\hat{q}_{ij}\psi)(u, o) = q_{ij}(u)\psi(u, o)$, since all involved operators commute and thus the othogonal matrices cancel, just as in the classical case. Therefore, the conjugate momenta $\hat{p}^{ij}$ to $\hat{q}_{ij}$ are given by the same formula as before and do not touch the gauge degrees of freedom at all.

The above construction accomplishes our goal of representing the vielbein fields in the quantum theory. However, two comments are in order.

First of all, it is generally desirable to replace $\mathrm{O}(n)$ by $\mathrm{SO}(n)$, because we want to confine to frames with a consistent orientation on all points in the manifold, or on all lattice sites respectively. This is because a well–defined $\mathrm{SO}(n)$ frame bundle is necessary to have an actual spin structure and therefore a well–defined notion of spinors on the manifold.

Second, we can replace $\mathrm{SO}(n)$ by its double cover $\mathrm{Spin}(n)$ and use the covering map $\pi : \mathrm{Spin}(n) \to \mathrm{SO}(n)$ to define the vielbein operators. This allows for the presence of additional spin degrees of freedom in the quantum theory.

An alternative Hilbert space is therefore given by the following definition.

**Definition 5.3** (Hilbert Space for the Vielbein Representation with Spin)**.** The Hilbert space for the vielbein representation is given by

$$\mathcal{H} = L^2(\mathrm{UT}_+(n, \mathbb{R}), \rho(u)\,\mathrm{d}u) \otimes L^2(\mathrm{Spin}(n), \mathrm{d}\nu), \tag{5.6}$$

where $\mathrm{d}\nu$ is the Haar measure on $\mathrm{Spin}(n)$. The vielbein variables are represented as multiplication operators $\hat{e}^i{}_a$ as follows:

$$(\hat{e}^i{}_a \psi)(u, s) = \sum_j (\pi(s))^i{}_j u_{ja} \psi(u, s). \tag{5.7}$$

Here, $\pi : \mathrm{Spin}(n) \to \mathrm{SO}(n)$ is the covering map.

Again, it is easy to see that the representation of the variables associated with the metric does not interact with the gauge degrees of freedom.

An interesting aspect of these representations is that, due to the tensor product structure, the gauge degrees of freedom are nicely separated. Gauge invariant operators don't interact with the pure gauge components of the quantum state at all. Therefore, they can be separated from the physical degrees of freedom by a partial trace.

## 6 Weyl Quantization

In this section, we introduce a generalization of the Weyl quantization formula (cf. Hall 2013; Weyl 1927) that is adapted to the non–standard representations constructed in this paper. This will allow us to quantize a large class of classical observables on our newly obtained Hilbert spaces. It will be of particular use in lattice quantum gravity, where dreaded square roots of $\det(q)$ are ubiquitous, which may be hard to quantize with conventional operator ordering techniques (cf. Lang and Schander 2023b).

Let $f(q, p) = (uq + vp)^n$, where $u, v \in \mathbb{R}$. Recall that the Weyl quantization formula in the standard representation is constructed in such a way as to ensure that the quantization scheme

$$Q[f] = (u\hat{q} + v\hat{p})^n \tag{6.1}$$

holds. Let us denote the Fourier transform of $f$ by $\tilde{f}$. A formula that realizes the above quantization prescription was given by Weyl:

$$Q[f] = \iint_{\mathbb{R}^2} \tilde{f}(\xi, \kappa)\, e^{i(\xi\hat{q} + \kappa\hat{p})}\, d\xi\, d\kappa. \tag{6.2}$$

The advantage of this formula is that it generalizes beyond mere polynomials and thus allows the quantization of a large class of classical phase space functions. By applying the identity

$$e^{i(\xi\hat{q} + \kappa\hat{p})} = e^{\frac{1}{2}i\xi\kappa} e^{i\xi\hat{q}} e^{i\kappa\hat{p}} \tag{6.3}$$

and applying the operator to a wave function $\psi$, one can perform further simplifications and arrive at the formula

$$(Q[f]\psi)(x) = \iint_{\mathbb{R}^2} f\left(\frac{x+y}{2}, p\right)\psi(y) e^{i(x-y)p}\, dp\, dy. \tag{6.4}$$

We want to devise a similar quantization scheme for the non–standard representations defined earlier. In order to do that, we would like to replace $e^{i\kappa\hat{p}}$ by the generalized shift operator $U(\kappa)$. The first obstacle we face is the fact that only $\kappa \in \mathbb{R}_+$ is allowed in $U(\kappa)$. We can use the classical identity $\tilde{f}(\kappa) = \tilde{f}(-\kappa)^*$, which holds for real valued functions $f$, in order to rewrite the integral over the negative real axis in the inverse Fourier transform as follows:

$$\int_{\mathbb{R}_-} \tilde{f}(\kappa)\, e^{i\kappa p}\, d\kappa = \int_{\mathbb{R}_-} \tilde{f}(-\kappa)^*\, e^{i\kappa p}\, d\kappa = \left(\int_{\mathbb{R}_-} \tilde{f}(-\kappa)\, e^{-i\kappa p}\, d\kappa\right)^* = \left(\int_{\mathbb{R}_+} \tilde{f}(\kappa)\, e^{i\kappa p}\, d\kappa\right)^*. \tag{6.5}$$

Therefore, the inverse Fourier transform for real valued $f$ can be rewritten as follows:

$$\int_{\mathbb{R}} \tilde{f}(\kappa)\, e^{i\kappa p}\, d\kappa = \int_{\mathbb{R}_+} \tilde{f}(\kappa)\, e^{i\kappa p}\, d\kappa + \left(\int_{\mathbb{R}_+} \tilde{f}(\kappa)\, e^{i\kappa p}\, d\kappa\right)^*. \tag{6.6}$$

Thus, only positive $\kappa$ appear in the integral. This trick will later allow us to replace the $e^{i\kappa p}$ factor by the generalized shift operator $U(\kappa)$.

Next, we want to generalize equation (6.3). We will do that by defining for $\xi \in \mathbb{R}$ and $\kappa, t \in \mathbb{R}_+$ the operators

$$U_{\xi,\kappa}(t) := e^{\frac{1}{2}it^2\xi\kappa} e^{it\xi\hat{p}} U(t\kappa) \tag{6.7}$$

and proving that they define a contraction semigroup whose generator is given by $\xi\hat{q} + \kappa\hat{p}$. This allows us to interpret $U_{\xi,\kappa}(1)$ as $e^{i(\xi\hat{q}+\kappa\hat{p})}$ by the Hille–Yosida theorem (theorem 2.3).

**Lemma 6.1.** *For all $\xi \in \mathbb{R}$ and $\kappa \in \mathbb{R}_+$, the set $\{U_{\xi,\kappa}(t) \mid t \in \mathbb{R}_+^n\}$ forms a contraction semigroup whose generator is given by $\xi\hat{q} + \kappa\hat{p}$.*

*Proof.* We need to check the properties in definition 2.1. It is trivial to see that $U_{\xi,\kappa}(0) = \mathrm{id}$. Moreover, strong continuity follows from the strong continuity of $U(\kappa)$ and the fact that the remaining factors are compositions of continuous maps. The semigroup property follows from a simple calculation:

$$\begin{aligned}
U(s\xi, s\kappa)U(t\xi, t\kappa) &= e^{is^2\frac{\xi\kappa}{2}} e^{is\xi\hat{x}} U(s\kappa) e^{it^2\frac{\xi\kappa}{2}} e^{it\xi\hat{x}} U(t\kappa) = e^{i(s^2+t^2)\frac{\xi\kappa}{2}} e^{is\xi\hat{x}} U(s\kappa) e^{it\xi\hat{x}} U(t\kappa) \\
&= e^{i(s^2+t^2)\frac{\xi\kappa}{2}} e^{is\xi\hat{x}} e^{it\xi(\hat{x}+s\kappa)} U(s\kappa) U(t\kappa) = e^{i(s+t)^2\frac{\xi\kappa}{2}} e^{i(s+t)\xi\hat{x}} U((s+t)\kappa) \\
&= U((s+t)\xi, (s+t)\kappa).
\end{aligned} \tag{6.8}$$

For the contraction property, we find that

$$\|U_{\xi,\kappa}(t)\| \le \|e^{it\xi\hat{q}}\| \|U(\kappa t)\| = \|U(\kappa t)\| \le 1, \tag{6.9}$$

where the last inequality is due to lemma 3.3. Thus, we are allowed to apply the Hille–Yosida theorem and compute the generator:

$$\left(\frac{\mathrm{d}}{\mathrm{d}t} U(t\xi, t\kappa)\right)_{t=0} = \left(\frac{\mathrm{d}}{\mathrm{d}t} e^{it^2\frac{\xi\kappa}{2}}\right)_{t=0} + \left(\frac{\mathrm{d}}{\mathrm{d}t} e^{it\xi\hat{x}}\right)_{t=0} + \left(\frac{\mathrm{d}}{\mathrm{d}t} U(t\kappa)\right)_{t=0} = \mathrm{i}(\xi\hat{x} + \kappa\hat{p}) \tag{6.10}$$

This completes the proof of lemma 6.1. □

This allows us to define the generalized Weyl quantization formula:

**Definition 6.2** (Generalized Weyl Quantization Formula). Let $f(q, p)$ be a real valued function on phase space. We define the generalized Weyl quantization of $f$ by

$$Q[f] = \bar{Q}[f] + \bar{Q}[f]^\dagger \tag{6.11}$$

where

$$\bar{Q}[f] = \frac{1}{(2\pi)^{2n}} \iint_{\mathbb{R}^n \times \mathbb{R}_+^n} \tilde{f}(\xi, \kappa)\, e^{\frac{1}{2}\mathrm{i}\xi\kappa} e^{\mathrm{i}\xi\hat{q}} U(\kappa)\, \mathrm{d}\xi\, \mathrm{d}\kappa. \tag{6.12}$$

We need to check that this provides a reasonable generalization of the original Weyl formula on polynomial functions. Since canonical pair commute among each other, we will just show this in one dimension. Let's assume that $f(q, p) = (uq + vp)^n$. We first need to compute its Fourier transform.

$$\begin{aligned}
\hat{f}(\xi, \kappa) &= \iint_{\mathbb{R}^2} f(q, p)\, e^{-i(\xi x + \kappa p)}\, dq\, dp = \iint_{\mathbb{R}^2} (uq + vp)^n\, e^{-i(\xi x + \kappa p)}\, dq\, dp \\
&= \frac{1}{2uv} \iint_{\mathbb{R}^2} s^n\, e^{-i\left(\xi \frac{s+t}{2u} + \kappa \frac{s-t}{2v}\right)}\, ds\, dt = \frac{1}{2uv} \iint_{\mathbb{R}^2} s^n\, e^{-is\left(\frac{\xi}{2u} + \frac{\kappa}{2v}\right)} e^{-it\left(\frac{\xi}{2u} - \frac{\kappa}{2v}\right)}\, ds\, dt \\
&= \frac{2\pi^2 i^n}{uv} \delta^{(n)}\!\left(\frac{\xi}{2u} + \frac{\kappa}{2v}\right) \delta\!\left(\frac{\xi}{2u} - \frac{\kappa}{2v}\right).
\end{aligned} \tag{6.13}$$

We insert this result into the formula for $\bar{Q}[f]$ without expanding $U(\xi, \kappa)$ and get

$$\begin{aligned}
\bar{Q}[f] &= \frac{1}{(2\pi)^2} \iint_{\mathbb{R} \times \mathbb{R}_+} \tilde{f}(\xi, \kappa)\, U(\xi, \kappa)\, d\xi\, d\kappa \\
&= \frac{i^n}{2uv} \iint_{\mathbb{R} \times \mathbb{R}_+} \delta^{(n)}\!\left(\frac{\xi}{2u} + \frac{\kappa}{2v}\right) \delta\!\left(\frac{\xi}{2u} - \frac{\kappa}{2v}\right) U(\xi, \kappa)\, d\xi\, d\kappa \\
&= i^n \iint_{\mathbb{R} \times \mathbb{R}_+} \delta^{(n)}(s)\delta(t)\, U(u(s+t), v(s-t))\, ds\, dt.
\end{aligned} \tag{6.14}$$

Using $((d^n/ds^n)U(su, sv))_{s=0} = i(u\hat{q} + v\hat{p})^n$, we can further simplify this to

$$\bar{Q}[f] = i^n (-1)^n \frac{1}{2} \left(\frac{d^n}{ds^n} U(su, sv)\right)_{s=0} = \frac{1}{2}(-i)^n (iu\hat{q} + iv\hat{p})^n = \frac{1}{2}(u\hat{q} + v\hat{p})^n. \tag{6.15}$$

Consequently, the generalized Weyl quantization of $f(q, p) = (uq + vp)^n$ is given by

$$Q[f] = \frac{1}{2}(u\hat{q} + v\hat{p})^n + \frac{1}{2}(u\hat{q} + v\hat{p}^\dagger)^n, \tag{6.16}$$

which appears to be the most natural extension of the Weyl quantization scheme to momentum operators that are not self–adjoint. If $\hat{p}$ is self–adjoint, this reduces to the standard formula.

## 7 Conclusion

In this paper, we have proposed a novel method to obtain (in general) non–standard representations of the canonical commutation relations. This is possible despite the celebrated Stone–von–Neumann theorem, because these representations generally do not exponentiate to the standard Weyl relations. While this may seem peculiar at first, we note that this already happens in textbook quantum mechanics. Even though it may not be widely known, the radial coordinate of the hydrogen atom and its conjugate momentum already have this property. Also in the case of the infinite potential well, this phenomenon occurs (cf. Hall 2013).

Our method is of particular use in situations where non–trivial configuration spaces need to be quantized and dependencies between the configuration variables need to be implemented in the algebra of observables. We were led to this proposal during the study of the canonical quantization of gravity, where the configuration variables are given by the matrix elements of the spatial metric, which, for physical reasons, needs to be represented in a manifestly symmetric and positive definite way. However, applications in many other areas of physics are conceivable as well. While one may often be able to arrive at such representations using brute–force calculations, our approach has the advantage to be systematic and widely applicable, which is especially useful in higher–dimensional configuration spaces. The complexity is completely reduced to the task of merely computing a Jacobian matrix.

The central issue of interest in our case is the representation of the matrix elements of a symmetric, positive definite matrix on a Hilbert space in such a way that these properties are maintained in the quantum theory. We achieve this by parameterizing the positive definite symmetric matrices in terms of their Cholesky decomposition. This decomposition induces a bijection between the positive definite symmetric matrices and the upper triangular matrices with positive diagonal elements. By applying our method to this map, we obtain a representation of the matrix elements $\hat{q}_{ab}$ of the spatial metric and their conjugate momenta $\hat{p}^{ab}$ with the additional constraint that $\hat{q}_{ab}s^a s^b$ is always a positive operator. This is possible without introducing additional gauge degrees of freedom. To the best of our knowledge, no analogous result can be found in the literature as of today.

In addition, we have obtained a representation of the vielbein fields $\hat{e}^i{}_a$ on the same Hilbert space with a clean separation of physical and gauge degrees of freedom in terms of a tensor product. Physical observables act trivially on the gauge degrees of freedom. Thus, passing to the gauge invariant Hilbert space is a simple as taking the partial trace with respect to the gauge degrees of freedom.

Moreover, we have generalized the Weyl quantization formula to the representations of the canonical commutation relations that arise using our method. This allows to quantize a large class of functions on the classical phase space on our representation Hilbert space. The resulting operators are automatically symmetric as in the standard representation and thus have the chance to possess self–adjoint extensions.

Our results are intended for applications in lattice quantum geometrodynamics, where each lattice site contributes a representation of the spatial metric and its conjugate momentum variables to the full tensor product Hilbert space of the lattice theory. We can thus proceed to quantize lattice discretized versions of the Hamiltonian and diffeomorphism constraints, as will be discussed in an upcoming paper of this series. This is the first step towards a fully non–perturbative approach to canonical quantum gravity and the study of the continuum limit. In contrast to LQG, our lattice discretization is not based on singularly smeared fields. As a consequence, it is possible to work on separable Hilbert spaces where states modeling similar geometries typically overlap instead of being orthogonal. In turn, strongly continuous representations of continuous symmetries like the diffeomorphism group are within reach. We will further examine these issues in a future paper.

*Acknowledgements*

Research at Perimeter Institute is supported in part by the Government of Canada through the Department of Innovation, Science and Economic Development and by the Province of Ontario through the Ministry of Colleges and Universities.